\documentclass[aip,apl,reprint,showpacs,groupedaddress]{revtex4-1}
\usepackage{amssymb}
\usepackage{graphicx}
\usepackage{amsmath}
\usepackage{graphicx}
\usepackage{dcolumn}
\usepackage{bm}
\usepackage{hyperref}
\usepackage[version=4]{mhchem}
\usepackage{xcolor} 

\usepackage{color}

\begin{document}

\setcounter{page}{1}

\title{Stability and electronic properties of planar defects in quaternary I$_2$-II-IV-VI$_4$ semiconductors}

\author{Ji-Sang Park}
\affiliation{Thomas Young Centre and Department of Materials, Imperial College London, Exhibition Road, London SW7 2AZ, UK}
\email[]{ji-sang.park@imperial.ac.uk}

\author{Sunghyun Kim}
\affiliation{Thomas Young Centre and Department of Materials, Imperial College London, Exhibition Road, London SW7 2AZ, UK}

\author{Aron Walsh}
\affiliation{Thomas Young Centre and Department of Materials, Imperial College London, Exhibition Road, London SW7 2AZ, UK}
\affiliation{Department of Materials Science and Engineering, Yonsei University, Seoul 03722, Korea}

\bibliographystyle{apsrev4-1}

\date{\today}
\begin{abstract}
Extended defects such as stacking faults and anti-site domain boundaries can perturb the band edges in Cu$_2$ZnSnS$_4$ and Cu$_2$ZnSnSe$_4$, acting as a weak electron barrier or a source for electron capture, respectively.
In order to find ways to prohibit the formation of planar defects, we investigated the effect of chemical substitution on the stability of the intrinsic stacking fault and metastable polytypes and analyze their electrical properties.
Substitution of Ag for Cu makes stacking faults less stable, whereas the other substitutions (Cd and Ge) promote their formation.
Ge substitution has no effect on the electron barrier of the intrinsic stacking fault, but Cd substitution reduces the barrier energy and Ag substitution makes the stacking fault electron capture.
While Cd substitution stabilizes the stannite structure, chemical substitutions make the primitive-mixed CuAu (PMCA) structure less stable with respect to the ground-state kesterite structure.
\end{abstract}

\maketitle

Cu$_2$ZnSn(S,Se)$_4$ (CZTSSe) has attracted much attention as it has material properties suitable for photovoltaic applications.\cite{polizzotti2013state,walsh2012kesterite,wallace2017steady,yan2018cu} 
The tunable direct band gap (1.0$\sim$1.5 eV) and the resulting high absorption coefficient\cite{chen2009crystal} make the material ideal to achieve high solar conversion efficiency.\cite{shockley1961detailed}
The record solar conversion efficiency, however, is lower than other mature technologies mainly because of the low open-circuit voltage,\cite{AENM201301465} 
which is thought to mainly originate from short minority carrier lifetime and presence of electronic band tails.\cite{grenet2018analysis}

Recently, the research field has been further extended to other materials with the kesterite structure to achieve a better device.
Many efforts have been devoted to replacing each component of the host material with other chemical elements in the same column in the periodic table.\cite{yuan2015engineering,collord2015combinatorial,choi2015electronic,kim2016improvement,gershon2016photovoltaic,collord2016germanium,Su2016,Guchhait2016,Marquez2017,yan2017beyond,tay2018solution}
Change in the fundamental material properties like the lattice constants and the band edge positions affect the defect properties, and thus the effect of chemical element substitution on the defects should be thoroughly investigated.

\begin{figure}
\includegraphics[width=\columnwidth]{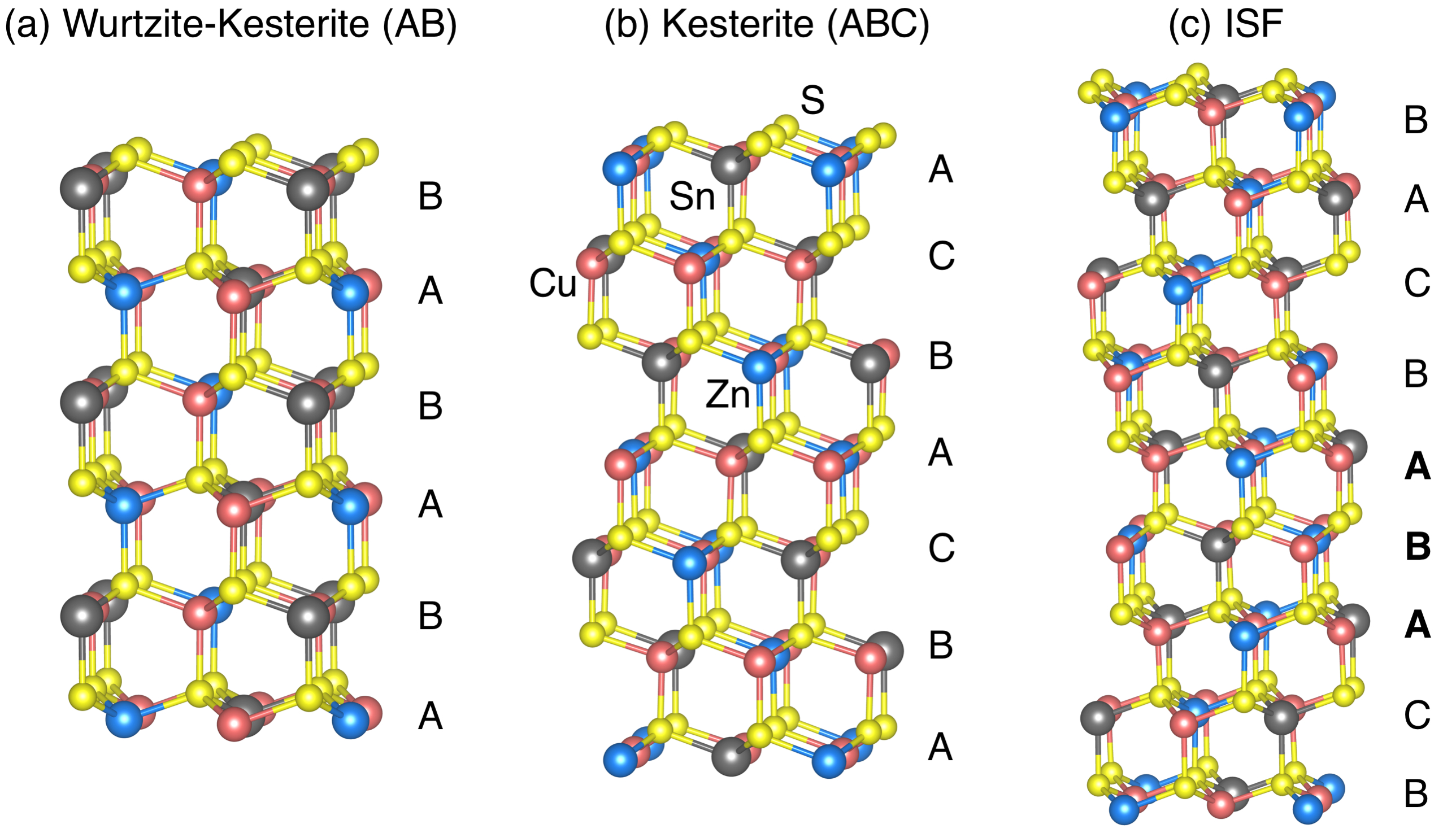}
\caption{\label{fig. 1} Atomic structure of Cu$_2$ZnSnS$_4$ with different stacking sequences. 
An internal stacking fault can be generated by removing a single layer from the perfect kesterite structure.
The corresponding polymorphs of CuInS$_2$ (a chalcopyrite) can be generated by substituting Zn and Sn by In. 
}
\end{figure}

Point defects and disorder in the kesterite materials have been investigated through many studies.\cite{walsh2012kesterite,yuan2015engineering,tiwari2017spectroscopic,kim2018identification,Wallace2018,park2018point} 
Studies of extended defects in this material, on the other hand, have been gaining attention recently.\cite{kattan2015crystal,C6NR04185J,park2018opposing}
Generally speaking, the extended defects perturb the band edge position of a host material because of the broken translational symmetry.\cite{stampfl1998energetics,Yan2001,yan2004energetics,liu2005luminescence,Sun2013} 
For example, in materials that are stable in the zinc-blende (ZB) structure (e.g. CdTe), a stacking fault (SF) can be understood as if the wurtzite (WZ) phase is formed locally.\cite{Yan2001} 
An experiment study reported the significantly hampered charge conduction by the planar defect in CdTe even though the conduction band offset between ZB CdTe and WZ CdTe is only 65 meV.\cite{Sun2013}
Similarly, we have found that SFs can be easily formed in Cu$_2$ZnSnS$_4$ and Cu$_2$ZnSnSe$_4$ and weakly inhibit electron transport.\cite{park2018opposing}
Existence of twin boundaries in this category of materials is also experimentally verified.\cite{kattan2015crystal}
Another example is anti-site domain boundaries (ADB), which are formed by arrays of anti-site defects.\cite{C6NR04185J} When an ADB is formed in CZTSSe by the shift of a Cu-Sn layer by ($a$/2,$a$/2,0), then a primitive-mixed CuAu (PMCA) phase is locally formed and the planar defect acts as a weak electron capture.\cite{park2015ordering,park2018opposing} 
The effect of extended defects can be minimized either by suppressing their formation or by reducing their perturbation on the underlying electronic structure.

\begin{table*}[ht]
\begin{center}
\caption{ 
Calculated physical properties of I$_2$-II-IV-VI$_4$ and I-III-VI$_2$ materials, the formation energy ($E_{f}$) of an intrinsic stacking fault (ISF), and effect of the defect on the band edges in each material. The average effective masses of bulk were calculated using the harmonic mean in units of $m_e$.}
\begin{tabular}{llccccccc}
\hline \hline 
Material     & $a$ ({\AA}) & $c/2$ ({\AA}) & E$_{g}$ (eV) & $\overline{m}_e$ & $\overline{m}_h$  & $E_{f}$(ISF) (eV/nm$^2$) & VBO (meV) & CBO (meV) \\ \hline \hline
Cu$_2$ZnSnS$_4$ (CZTS)  & 5.441 & 5.425 &  1.46 & 0.18 & 0.40& 0.14 & 6 & 28 \\
Cu$_2$ZnSnSe$_4$ (CZTSe) & 5.722 & 5.702  & 0.89 & 0.10 & 0.23  & 0.18 & 9 & 29 \\
\hline
Ag$_2$ZnSnS$_4 $ (AZTS) & 5.832 & 5.472 & 1.72 & 0.18 & 0.46 & 0.26 & -50 & -19 \\	
Ag$_2$ZnSnSe$_4$ (AZTSe) & 6.089 & 5.753 & 1.05 &  0.11 & 0.26 & 0.27 & -42 & -12 \\	 \hline
Cu$_2$CdSnS$_4 $ (CCTS) & 5.537 & 5.599 & 1.27 & 0.16  & 0.88 & -0.04 & 8 & 11 \\	
Cu$_2$CdSnSe$_4$ (CCTSe) & 5.822 & 5.869 & 0.73 &  0.08 & 0.71 & 0.05 & 15 & 23 \\	\hline
Cu$_2$ZnGeS$_4 $ (CZGS) & 5.347 & 5.282 &  2.06 & 0.21 & 0.63 & 0.08 & 3 & 30 \\	
Cu$_2$ZnGeSe$_4$ (CZGSe) & 5.631 & 5.573 & 1.18 & 0.12 & 0.27 & 0.07 & -8 & 30  \\	\hline
CuInSe$_2$ (CISe)     & 5.811 & 5.870 & 1.11 & 0.01 & 0.22 & 0.16 & 25 & 30 \\
CuGaSe$_2$ (CGSe)   & 5.543 & 5.600 & 1.77 & 0.13 & 0.32 & 0.12 & 15 & 49 \\	\hline \hline 
\end{tabular}

\end{center}
\end{table*} 

One popular way to change the material properties is to replace chemical constituents with isovalent elements. 
Kesterite has ABC stacking sequence along the [112] direction, but a wurtzite-derived polytype of kesterite (i.e. wurtzite-kesterite) has the AB stacking sequence along the [0001] direction, as shown in Figure 1.
Therefore, the formation of a SF in kesterite can be understood as if thin wurtzite-kesterite is locally formed between two kesterite grains.
Chen \textit{et al.} previously performed density functional theory (DFT) calculations using generalized gradient approximation (GGA) and claimed that wurtzite-kesterite becomes less unstable if Cu, Zn, and Sn in CZTS are replaced with Ag, Cd, and Ge, respectively.\cite{chen2010wurtzite} 
Based on their conclusion, one can expect that the formation of SF defects in the absorber material will be promoted by the substitutions.
Coincidentally, these elements have been suggested to improve the device efficiency.\cite{yuan2015engineering,gershon2016photovoltaic,Su2016,Guchhait2016,Marquez2017,yan2017beyond} 
This circumstance has motivated us to examine the effects of elemental substitution on the physical properties of intrinsic stacking fault (ISF).

We investigated the effect of cation substitutions in CZTSSe on the stability and electronic properties of extended defects.
Stability of other polytypes were also investigated to understand the effect of alloying on the formation of anti-site domain boundaries.
The electronic band gap and effective masses of each material were also obtained.

We performed first-principles density functional theory (DFT) calculations to investigate the SFs in multi-cation quaternary I$_2$-II-IV-VI$_4$ semiconductors.
We used the hybrid functional proposed by Heyd, Scuseria, and Ernzerhof\cite{HSE} and the projector augmented wave (PAW) pseudo-potentials,\cite{PAW} as implemented in the VASP code.\cite{PhysRevB.54.11169}
The screening parameter and the exchange parameter were set to 0.2 {\AA}$^{-1}$ and $\alpha$ = 0.25 (0.3 for CuIn(Ga)Se$_2$),\cite{park2014defect} respectively.
The lattice parameters and the internal coordinates were fully relaxed until the residual force becomes smaller than 0.03 eV {\AA}$^{-1}$.
The wavefunctions are expanded in plane waves with an energy cutoff of 400 eV.
For Brillouin zone (BZ) integration, the smallest spacing between \textit{k}-points was set to $\simeq$0.05 {\AA}$^{-1}$.

The formation energy of an ISF and its effect on the band edges were calculated as explained in the previous paper.\cite{park2018opposing}
For each material, the valence band offset (VBO) and the conduction band offset (CBO) were calculated by the difference between the band edges of a supercell with an ISF and those of bulk. 
The electrostatic potential at far from the SF was used as a reference to estimate the band edges of bulk.\cite{park2018quick}
Each supercell with an ISF contains 8 double-layers in which a single layer have 4 atoms.
We note that the kesterite materials feature a direct band gap at the $\Gamma$ point.

The calculated formation energy ($E_f$) of an ISF defect and the band edge positions with respect to the bulk counterparts are summarized in Table 1.
The SFs are more likely formed in CZGS(Se) as compared to CZTS(Se) in our hybrid DFT calculations, consistent with the fact that the wurtzite-kesterite becomes less unstable by the Ge incorporation.\cite{chen2010wurtzite}
However, substitution of Sn by Ge does not affect the barrier energy of an ISF, and thus we conclude that the effect of SFs can be more prominent in CZGS(Se) if the grain size is similar.
However, the SFs could be less detrimental in CZTS with a small amount of Ge as Ge substitution seems promote the crystallization (larger grains).\cite{Giraldo2018}

Cd does not contribute to the band edge electronic structure, so their effects should be understood in terms of the strain.
In this regard, it is worth mentioning how the strain affects the relative stability of wurtzite and zinc-blende polymorphs. 
A simple formula has been proposed for calculating the energy difference between zinc-blende and wurtzite,\cite{ito1998simple} 
where the electrostatic interaction between the covalent bond charge (located at the center of the bond) stabilizes zinc-blende. 
On the other hand, there is an attractive interaction between positive and negative charges at the atomic sites (i.e. cations and anions). 
A material is stable in zinc-blende or a zinc-blende-derived structure if the repulsive interaction of covalent bond charges is greater than the attractive interaction.
Under a hydrostatic tensile strain, the repulsive interaction is weakened more as both interactions are proportional to the inverse of the distance, making the wurtzite phase less unstable. 

\begin{figure}
\includegraphics[width=\columnwidth]{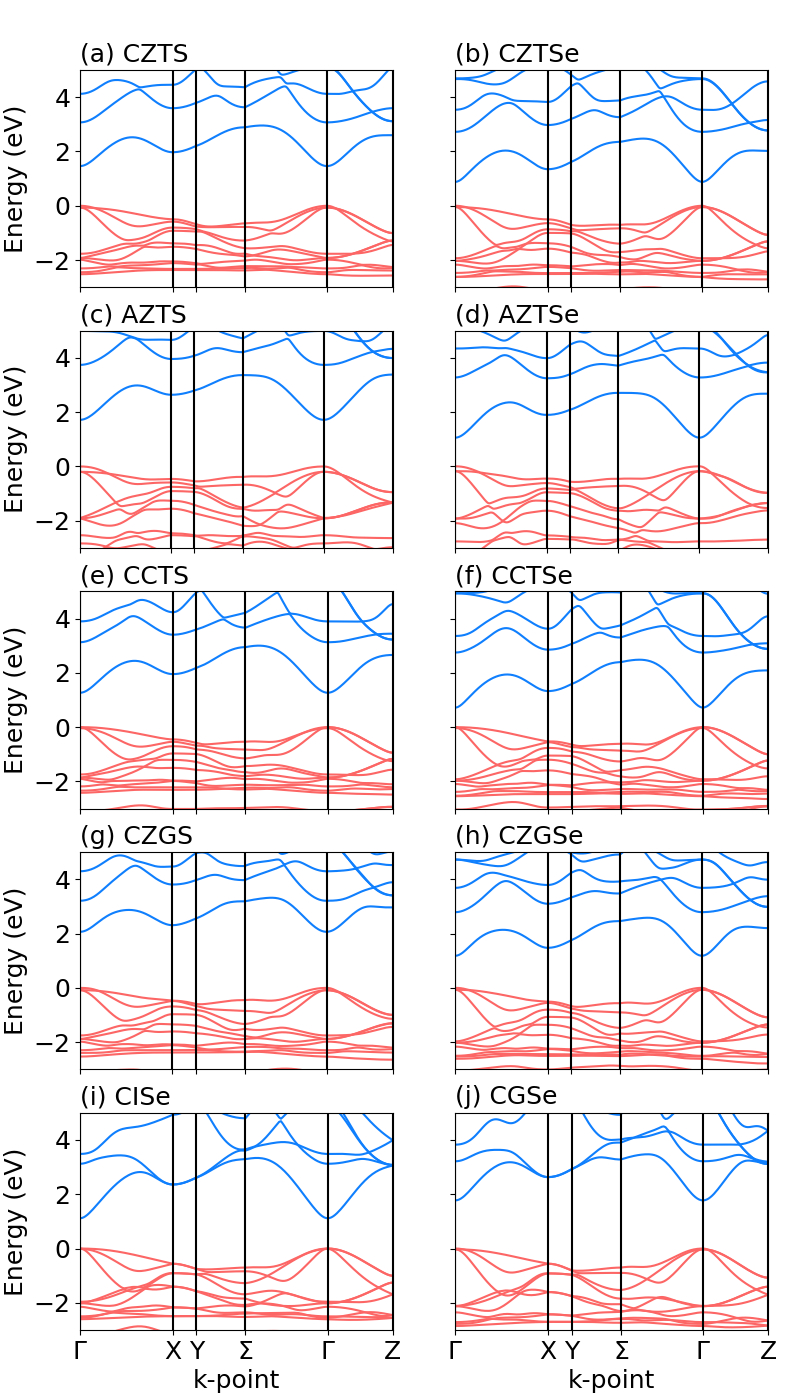}
\caption{\label{fig. 2} Electronic band structure of the kesterite materials obtained from the HSE06/DFT electronic structure using Wannier90. The upper valence band of each material is set to 0 eV. Vertical lines represent the special \textit{k}-points which are noted in the bottom sub-figures.
}
\end{figure}

Consistent with the prediction, the formation energy of the ISF in Cu$_2$CdSnS$_4$ (CCTS) and Cu$_2$CdSnSe$_4$ (CCTSe) (Table 1) is lower than that in CZTS(Se) by more than 10 meV/nm$^2$.
The formation energy in CCTS is calculated to be negative because CCTS is stabilized in wurtzite-kesterite in our HSE06/DFT calculation.
An ISF in CCTS(Se) becomes a slightly weaker electron barrier as compared to CZTS(Se).
If Cd atoms are provided from CdS and thus some Zn atoms are replaced by Cd, 
then it will promote the formation of SFs in the vicinity of the CZTS/CdS interface because of the increased lattice constants.
A recent experimental study also reports the elemental intermixing of Cd and Zn at CZTS/CdS interface.\cite{yan2018cu}

Ag substitution has been suggested as a potential strategy to reduce the Cu-Zn disorder and increase the open-circuit voltage.\cite{yuan2015engineering,chagarov2016ag2znsn}
Even though wurtzite exhibits a higher conduction band than zinc-blende in every material considered in a previous study,\cite{murayama1994chemical} 
the band offsets are calculated to be negative, meaning that the SFs trap electron carriers.
To check whether this is reproduced using another exchange-correlation functional, we also calculated the offsets in AZTS using the SCAN exchange-correlation function\cite{SCAN} by applying Hubbard U on Ag $d$ and Zn $d$ by 6 eV. The atomic coordinates were fixed throughout the calculations. The VBO and the CBO were calculated to be $-42$ meV and $-20$ meV, respectively, consistent with the HSE06 results.
Even though the stacking faults in AZTS(Se) can be detrimental, stacking faults are expected to form less in AZTS(Se) because of the higher formation energies.

The physical properties of CZTS are often compared to CIGS,\cite{Shafarman2011} thus we also investigated an ISF in CuInSe$_2$ (CISe) and CuGaSe$_2$ (CGSe).
If the grain size is similar, SFs are slightly more favored in the chalcopyrites than CZTSe because of the lower formation energy.
An ISF in CISe raises the CBM by the similar amount to that in CZTS, while that in CGSe increase the conduction band more.

We also obtained the HSE06 band structure and effective masses of the kesterite materials, which are summarized in Figure 3 and Table 1, respectively.
The electronic band structure and the effective masses were obtained by using Wannier90\cite{mostofi2014updated} and Effective Mass Calculator.\cite{fonari2012effective}
For CZTS, we obtained similar effective masses to a previous hybrid DFT study.\cite{liu2012first}

All substitutions make hole effective mass heavier, while Cd substitution results in a slightly lighter electron mass.
However, all kesterite materials exhibit heavier electron and hole masses than CuInSe$_2$, as illustrated in the band structures (Figure 3) and Table 1.
We also note that degeneracy of the conduction bands at the Z point is found in CuInSe$_2$ and CuGaSe$_2$, however, the bands are split in kesterites because of the segregating states formed by symmetry breaking.\cite{park2014defect,park2015ordering}

\begin{table*}[ht]
\begin{center}
\caption{ Relative formation energy ($E_f$), lattice constants ($a$ and $c$), the electronic band gap ($E_g$), and position of the conduction band minimum ($k_{\mathrm{CBM}}$) of stannite (ST) and the primitive-mixed CuAu (PMCA). $E_f$ of kesterite phase is set to 0. $\Gamma$ = $[0,0,0]$ and Z = $\frac 12(b_1 + b_2 + b_3)$ where $b_i$ stands for a reciprocal vector.}
\begin{tabular}{cccccccc}
\hline \hline 
Material  & polytype & $E_f$ (meV/atom)                    & $a$ (\AA)         & $c/2$ (\AA)          & $E_g$ (eV) & $k_{\mathrm{CBM}}$  & Ref. \\ \hline \hline
CZTS     & ST              &  4                          &  5.432         & 5.454          &  1.29    & $\Gamma$ & \cite{park2015ordering}  \\  
              & PMCA         &  5                          & 5.435        &  5.449        &  1.22            & $\Gamma$ & \cite{park2015ordering}  \\  \hline
CZTSe   & ST              &  5                          &  5.694       &  5.738         &   0.71          & $\Gamma$ & \cite{park2015ordering}  \\  
              & PMCA         &  7                          &  5.709       & 5.734         &   0.58           & Z  & \cite{park2015ordering}  \\   \hline
AZTS     & ST              &  19                        &  5.506        & 6.117          &  1.43           & $\Gamma$    \\  
              & PMCA         &  20                        &  5.505       & 6.119          &   1.42          &  $\Gamma$  \\  \hline
AZTSe   & ST              &  21                        &  5.797       & 6.331          &   0.72           &  $\Gamma$   \\
              & PMCA         &  22                        & 5.798        & 6.335         &   0.69           &  $\Gamma$   \\   \hline

CCTS     & ST              &  -5                        &  5.618        & 5.453       &   1.22            & $\Gamma$   \\  
              & PMCA         &  6                         & 5.605       & 5.475       &  1.12               & $\Gamma$    \\  \hline
CCTSe   & ST              &  -5                        & 5.882      &  5.741       &  0.69               &  $\Gamma$  \\
              & PMCA         &  7                         & 5.868      &  5.775       &  0.53               & Z    \\   \hline
CZGSe   & ST              &  7                         &  5.580     &  5.643           &   0.95        & $\Gamma$ &  \cite{choi2015electronic}  \\
              & PMCA         &  19                        &  5.573     & 5.643            &  0.36         & Z  & \cite{choi2015electronic}  \\   \hline \hline
\end{tabular}
\end{center}
\end{table*} 

Another type of extended defect formed in this category of materials is the ADB. Whether the ADB satisfy the octet rule depends on the fault displacement.\cite{C6NR04185J,park2018opposing} 
Stability of the defect satisfying the octet rule can be estimated by knowing how much the PMCA polymorph is unstable as compared to the kesterite phase. 
In our calculations, the Ag substitution makes the PMCA phase unstable with respect to the kesterite phase, as summarized in Table 2. 
Therefore, the Ag substitution inhibits the formation of the ADB with the fault displacement of an entire layer with cations by $\frac 12$[$a$,$a$,0].

\begin{figure}
\includegraphics[width=\columnwidth]{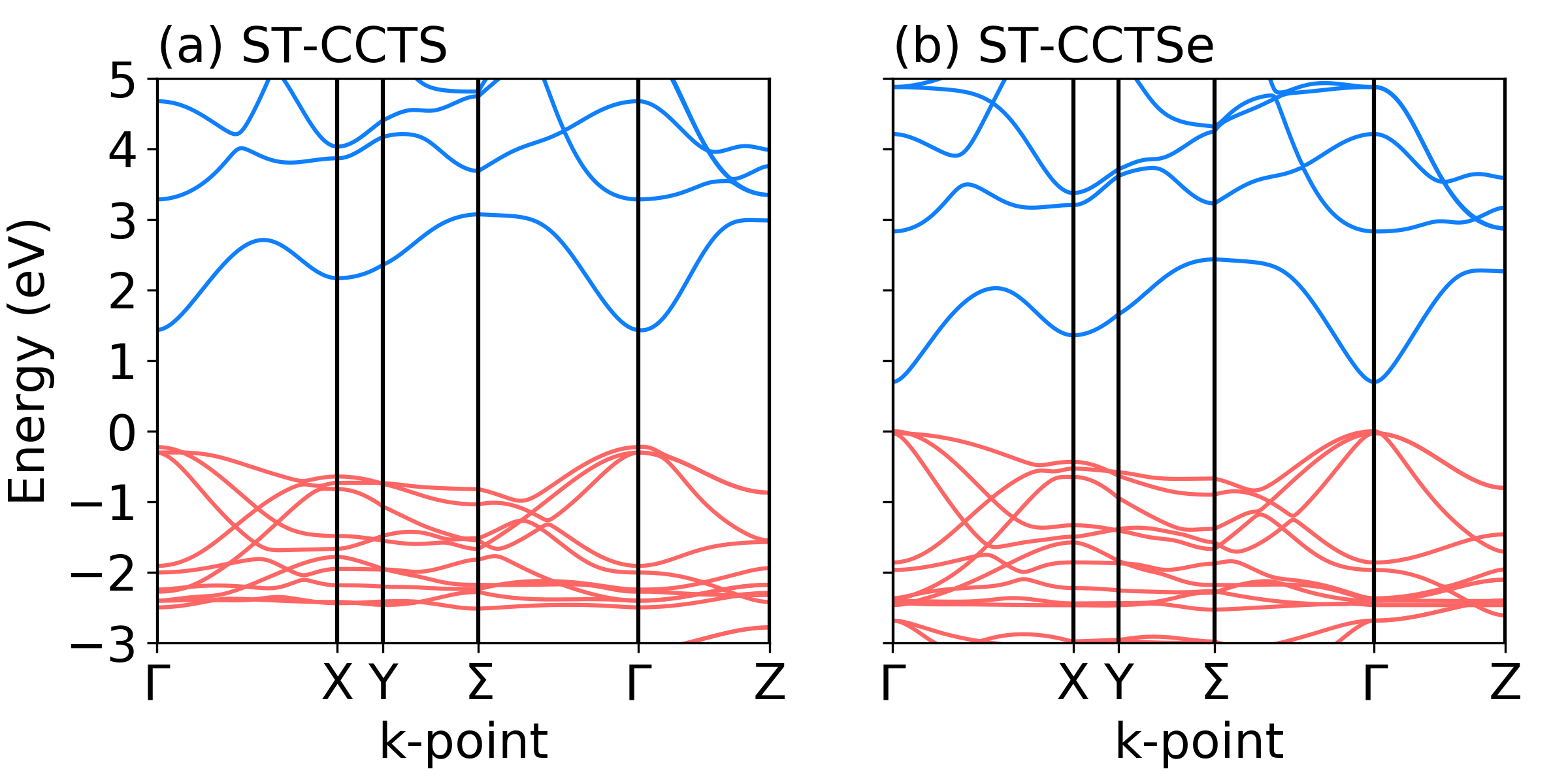}
\caption{\label{fig. 3} Electronic structure of stannite CCTS and stannite CCTSe calculated from HSE06/DFT. The VBM of each material is set to 0 eV. Vertical lines represent the special \textit{k}-points.
}
\end{figure}

Cd substitution doesn't affect the relative formation energy of PMCA with respect to kesterite much, but we expect that formation of the ADB is inhibited when Cd composition ratio is high as the stannite (ST) phase becomes more stable than kesterite, consistent with experimental findings.\cite{su2015cation} 
Since stannite and kesterite have similar band gap energy, thus coexistence of two phases will not result in strong band gap fluctuations.
As is found in CZTS(Se),\cite{park2015ordering} the band gap of ST and PMCA is generally lower than that of the kesterite phase.
The electronic band structure of stannite CCTS and CCTSe is shown in Figure 3. 
Stannite CCTS (CCTSe) has a similar electron effective mass of 0.15 (0.08) $m_e$, but a lighter hole mass of 0.34 (0.20) $m_e$ as compared to kesterite counterparts.

In summary, we examined the effect of chemical substitution on the stability and the electronic properties of ISF and ADB.
Although Ag substitution makes SFs an electron captures, it will help to increase the efficiency as less planar defects are formed. Substitution of Cd and Ge promotes the formation of SFs while the electron barriers are not lowered much, potentially resulting in the lower charge extraction unless the crystallinity is enhanced.

\begin{acknowledgments}
We thank Gilles Dennler for his constructive suggestions.
JP thanks the Royal Society for a Shooter International Fellowship.
This project has received funding from the European H2020 Framework Programme for research, technological development and demonstration under grant agreement no. 720907. See http://www.starcell.eu. 
Via our membership of the UK's HPC Materials Chemistry Consortium, which is funded by EPSRC (EP/L000202), this work used the ARCHER UK National Supercomputing Service (http://www.archer.ac.uk). 
We are grateful to the UK Materials and Molecular Modelling Hub for computational resources, which is partially funded by EPSRC (EP/P020194/1). 
\end{acknowledgments} 

\bibliography{czts}

\end{document}